\documentclass{article}
\usepackage{amsmath}

\begin{document}

\title{Manipulating Synchronous Optical Signals with a Double $\Lambda $
Atomic Ensemble}
\author{Zhuan Li$^{1}$, De-Zhong Cao$^{1}$, and Kaige Wang$^{2,1}\thanks{%
Corresponding author: wangkg@bnu.edu.cn}$ \\
1.Department of Physics, Applied Optics Beijing Area Major Laboratory,\\
Beijing Normal University, Beijing 100875, China\\
2.CCAST (World Laboratory), P. O. Box 8730, Beijing 100080, China}
\maketitle

\begin{abstract}
We analyze a double $\Lambda $ atomic configuration interacting with two
signal beams and two control beams. Because of the quantum interference
between the two $\Lambda $ channels, the four fields are phase-matched in
electromagnetically induced transparency. Our numerical simulation shows
that this system is able to manipulate synchronous optical signals, such as
generation of optical twin signals, data correction, signal transfer and
amplification in the atomic storage.

\end{abstract}

In modern communication, the transmission of information is usually carried
out among multi-users which are organized in a network. The network consists
of spatially separated nodes in which information can be stored and locally
manipulated. Recently, experiments have demonstrated that optical signal can
be stored in and then retrieved from an atomic ensemble\cite{liu}\cite{phili}%
. Light storage is implemented in a scheme of electromagnetically induced
transparency (EIT) in which the atoms with a $\Lambda $ configuration
interact resonantly with both a signal beam and a control beam\cite{harris}.
Theoretically, it has been proved that the scheme can also be used as a
quantum memory\cite{flei}-\cite{kaige}. Then, EIT is extended to a double $%
\Lambda $ configuration in which two couples of probe and control beams
interact resonantly with a four-level atom\cite{ari}-\cite{rac}. In this
model the two $\Lambda $ subsystems share a common dark state and the
quantum interference exists not only between the two lower states, but also
between the two $\Lambda $ channels. Therefore, the EIT effect occurs only
when the ratio of Rabi frequencies in each $\Lambda $ channel is equal. Ref.%
\cite{ari} pointed out the lossless propagation of shape matched probe
pulses interacting with one $\Lambda $ channel when the strong identical
pulses drive another $\Lambda $ channel. However, the storage mechanism of
optical pulse in the standard EIT interaction can be applied to double $%
\Lambda $ configuration where two probe pulses can be simultaneously stored
and released\cite{rac}. Considering the special features of double $\Lambda $
configuration, in this paper, we show that this atomic system is capable of
manipulating synchronous optical signals, such as generation of optical twin
signals, data correction, signal transfer and amplification.

We\ consider a double-$\Lambda $ configuration with two lower levels $%
\left\vert b\right\rangle $ and $\left\vert c\right\rangle ,$ and two upper
levels $\left\vert a\right\rangle $ and $\left\vert d\right\rangle $, as
shown in Fig. 1. The two weak signal fields $\Omega _{1s}$ and $\Omega _{2s}$
couple the atomic transitions $\left\vert b\right\rangle -\left\vert
a\right\rangle $ and $\left\vert b\right\rangle -\left\vert d\right\rangle $%
, respectively, while the two control fields $\Omega _{1c}$ and $\Omega
_{2c} $ couple the transitions $\left\vert c\right\rangle -\left\vert
a\right\rangle $ and $\left\vert c\right\rangle -\left\vert d\right\rangle $%
, respectively. We assume full resonance between fields and atomic
transitions, and the four levels of atom form a closed loop. The collective
atomic operators obeys the Heisenberg-Langevin equation\cite{kaige1}.
However, the propagation equations of the two signal beams are written as 
\begin{subequations}
\begin{eqnarray}
c\frac{\partial \Omega _{1s}}{\partial z}+\frac{\partial \Omega _{1s}}{%
\partial t} &=&\frac{i}{2}g_{1}^{2}N\rho _{ba},  \label{fa} \\
c\frac{\partial \Omega _{2s}}{\partial z}+\frac{\partial \Omega _{2s}}{%
\partial t} &=&\frac{i}{2}g_{2}^{2}N\rho _{bd},  \label{fb}
\end{eqnarray}%
where $g_{i}$ is the coupling coefficient between the signal beam and the
atomic transition, and $N$ is the number of atoms. In order to minimize the
parameters of the model, we assume a uniform atomic decay $\gamma $ and $%
g_{1}=g_{2}=g$.

Assuming that the atomic relaxation is much faster than the variation of the
signals, we may obtain approximately the autonomous equations for the two
plane-wave signals

\end{subequations}
\begin{subequations}
\label{s}
\begin{eqnarray}
\frac{\partial \Omega _{1s}}{\partial \tau } &=&-\eta \lbrack \frac{%
\left\vert \Omega _{2c}\right\vert ^{2}}{\left\vert \Omega _{1c}\right\vert
^{2}+\left\vert \Omega _{2c}\right\vert ^{2}}\Omega _{1s}-\frac{\Omega
_{1c}\Omega _{2c}^{\star }}{\left\vert \Omega _{1c}\right\vert
^{2}+\left\vert \Omega _{2c}\right\vert ^{2}}\Omega _{2s}],  \label{sa} \\
\frac{\partial \Omega _{2s}}{\partial \tau } &=&-\eta \lbrack \frac{%
\left\vert \Omega _{1c}\right\vert ^{2}}{\left\vert \Omega _{1c}\right\vert
^{2}+\left\vert \Omega _{2c}\right\vert ^{2}}\Omega _{2s}-\frac{\Omega
_{1c}^{\star }\Omega _{2c}}{\left\vert \Omega _{1c}\right\vert
^{2}+\left\vert \Omega _{2c}\right\vert ^{2}}\Omega _{1s}],  \label{sb}
\end{eqnarray}%
where $\eta =g^{2}N/(4\gamma ^{2})$ and $\tau =\gamma t$. Given the initial
values, Equation (\ref{s}) can be analytically solved. Then we set $\tau
\rightarrow \infty $ and obtain

\end{subequations}
\begin{subequations}
\label{6}
\begin{eqnarray}
\frac{\Omega _{1s}(\infty )}{\Omega _{1s}(0)} &=&\frac{1+\sqrt{\xi \mu }\exp
(i\delta _{0})}{1+\xi },  \label{6a} \\
\frac{\Omega _{2s}(\infty )}{\Omega _{2s}(0)} &=&\frac{\xi +\sqrt{\xi /\mu }%
\exp (-i\delta _{0})}{1+\xi }.  \label{6b}
\end{eqnarray}%
where $\xi =\left\vert \Omega _{2c}\right\vert ^{2}/\left\vert \Omega
_{1c}\right\vert ^{2},$ $\mu =\left\vert \Omega _{2s}(0)\right\vert
^{2}/\left\vert \Omega _{1s}(0)\right\vert ^{2}$, and $\delta _{0}=\arg
[\Omega _{1c}]-\arg [\Omega _{2c}]+\arg [\Omega _{2s}(0)]-\arg [\Omega
_{1s}(0)]$. By eliminating the initial values $\Omega _{1s}(0)$ and $\Omega
_{2s}(0)$, the four fields satisfy the relation 
\end{subequations}
\begin{equation}
\frac{\Omega _{2s}(\infty )}{\Omega _{1s}(\infty )}=\frac{\Omega _{2c}}{%
\Omega _{1c}},  \label{2cg}
\end{equation}%
under which the double-$\Lambda $ atom is in a dark state and hence the dual
EIT effect occurs\cite{ko}. Equation (\ref{2cg}) indicates the phase
matching relation $\delta \equiv \arg [\Omega _{1c}]-\arg [\Omega
_{2c}]+\arg [\Omega _{2s}(\infty )]-\arg [\Omega _{1s}(\infty )]=0$, and
implies shape-matched propagation of the two signal fields. Furthermore,
according to Eqs. (\ref{6}), the magnitudes of the transmitted signal fields
depend on their initial values, the intensity ratio of the control fields,
and the phase $\delta _{0}$. Figure 2 shows the amplification/attenuation
ratio of the signal intensity $\left\vert \Omega _{i}(\infty )\right\vert
^{2}/\left\vert \Omega _{i}(0)\right\vert ^{2}$ as a function of the
intensity ratio $\xi $ of the control fields for the three phases $\delta
_{0}=0,\pi /2$ and $\pi ,$ where the two signal beams at the initial time
have the same intensity ($\mu =1$). It can be seen from Fig. 2 that the
initial phase matching parameter $\delta _{0}$ dominates the amplification
of signals. For $\delta _{0}=0$, one of two signal beams can be amplified
(with the maximum amplification 1.457), while the phase matching maintains,
i.e. $\delta =\delta _{0}=0$. At $\xi =1$, the two signal beams, which are
equal initially, propagate transparently in the medium and the EIT\ effect
occurs. However, if $\delta _{0}=\pi $ is set, the two signal beams quench
simultaneously at $\xi =1$. This is so called electromagnetically induced
absorption (EIA)\cite{tai}.

In the following, we discuss the schemes using the double $\Lambda $ model
to manipulate synchronous signals.

(i) \emph{Generation of optical twin signals}. If only one signal enters
into the atomic medium, the modulation of the signal can be duplicated to a
new beam which may have different frequency and polarization. This effect
refers to a four-wave mixing (FWM) mechanics\cite{wu} but it contains
strongly quantum interference resulting in beam matching. In the analytical
solution (\ref{6}), by setting $\Omega _{1s}(0)=0$ we obtain $\Omega
_{1s}(\infty )=\frac{\sqrt{\xi }}{1+\xi }\Omega _{2s}(0)\exp [i\arg (\Omega
_{1c})-i\arg (\Omega _{2c})]$ and $\Omega _{2s}(\infty )=\frac{\xi }{1+\xi }%
\Omega _{2s}(0)$. For example, if $\Omega _{2c}=\Omega _{1c}$, the two
signal beams are identical, $\Omega _{1s}(\infty )=\Omega _{2s}(\infty
)=(1/2)\Omega _{2s}(0)$.

The numerical simulation of this scheme is shown in Fig. 3, in which all the
atoms are initially at the ground level $\left\vert b\right\rangle $ (the
same for Figs. 4-7 as well). When the two control beams are set to be equal,
we obtain the twin signals after a few times of atomic relaxation by
inputting only one signal beam.

(ii) \emph{Data correction}. If there are two identical digital signals
generated, for example, by the above scheme, and a few pulses are lost in
transmission, they can be corrected by parallel comparison of these two
series of pulses. In Figs. 4a and 4b, we consider such two imperfect digital
signals as input and set two equal control beams in the double $\Lambda $
atoms. The numerical simulation shows that two identical series of pulses
are resumed, as shown in Figs. 4c and 4d, in which the blank pulses are
recovered. This correction method is simple and effective. Only when two
corresponding pulses of the two beams are lost simultaneously, they are
non-recoverable. But this happens with a negligible probability.

(iii) \emph{Signal amplification in transmission}. If there are two
synchronous signals available, one of the signals can be amplified in this\
system. The analytical solution (\ref{6}) provides the indication how to
choose the parameters for amplification. When two signals are identical ($%
\mu =1$), for instance, the maximum amplification ratio 1.457 occurs for $%
\delta _{0}=0$ and $\xi =3\pm 2\sqrt{2}$ (see Fig. 2). In Fig. 5, we prepare
two identical pulses at time $\tau _{0}$. By setting the parameter $\xi
=0.1716$, the signal $\Omega _{1s}$ is amplified with the maximum
amplification.

(iv) \emph{Signal transfer from one beam to another}. The dual EIT system is
able to transfer a signal from one beam to another which may have different
frequency and polarization. The transfer process is implemented through the
atomic storage. In Fig. 6, a pulse is initially carried by the signal beam 1
while the control beams 1 and 2 are set to a high and a low levels,
respectively. When the levels of the two control beams are gradually
exchanged, the pulse has been transferred from one signal beam to another.

(v) \emph{Signal amplification in atomic storage}. In the signal storage
scheme carried out by a three-level EIT model, loss occurs in the atomic
storing process because of the atomic decay. However, in the dual EIT model,
it is possible to amplify optical signal in the atomic storing process.
Similar to scheme (iii), we should prepare two synchronous signals as input.
Figure 7 shows the signal amplification through the atomic storage. In the
first stage when the two control beams have the same high strength, the dual
EIT condition is satisfied and the two signals propagate transparently in
the medium. When the two control beams are decreased to a low level in the
second stage, the two signals are simultaneously stored in the atomic
ensemble. In the final stage in which only one control beam is recovered to
the previous high level, one of the signal beams has been retrieved with the
amplification ratio 1.452 while the other one vanishes.

In summary, we show that the double $\Lambda $ configuration atomic system
possesses additional quantum interference between the two EIT channels and
can provide powerful application to manipulate optical synchronous signals.

This research was supported by the National Fundamental Research Program of
China Project No. 2001CB309310, and the National Natural Science Foundation
of China, Project No. 60278021 and No. 10174007.

Captions of Figures:

Fig. 1 Atomic levels of a double $\Lambda $ configuration.

Fig. 2 Intensity amplification/attenuation ratios of the two signal beams as
functions of the control field parameter $\xi $ for the three phases $\delta
_{0}=0,\pi /2$ and $\pi $. The two input signals at the initial time have
the same intensity ($\mu =1$).

Fig. 3 Generation of twin signals. The two control beams are equal $\Omega
_{1c}/(g\sqrt{N})=\Omega _{2c}/(g\sqrt{N})=12$. In the pulse propagation,
the normalized times ($\tau =\gamma t$) are set as $\tau _{0}=0,\tau
_{1}=1.5,\tau _{2}=3,\tau _{3}=4.5,\tau _{4}=9.5,\tau _{5}=14.5,\tau
_{6}=19.5,\tau _{7}=24.5,\tau _{8}=29.5$. In Figs. 3-7, the coupling
strength is taken as $g\sqrt{N}/\gamma =2.$

Fig. 4 An example of data correction in which (a) and (d) are the two
imperfect pulse signals as input. In the propagation, the lost pulses are
recovered. The control beams are the same as in Fig. 3.

Fig. 5 Signal amplification in transmission. In the pulse propagation, the
normalized times are set as $\tau _{0}=0,\tau _{1}=1.5,\tau _{2}=3,\tau
_{3}=10.5,\tau _{4}=18,\tau _{5}=25.5,\tau _{6}=33.$

Fig. 6 Signal transfer process. (a) Evolution of the signal beam 1 and (b)
Evolution of the signal beam 2. The insets show variations of the
corresponding control beams.

Fig. 7 Signal amplification in atomic storage. The descriptions are the same
as in Fig. 6. Note that the two control beams are equal until the time $%
\gamma t=150$.

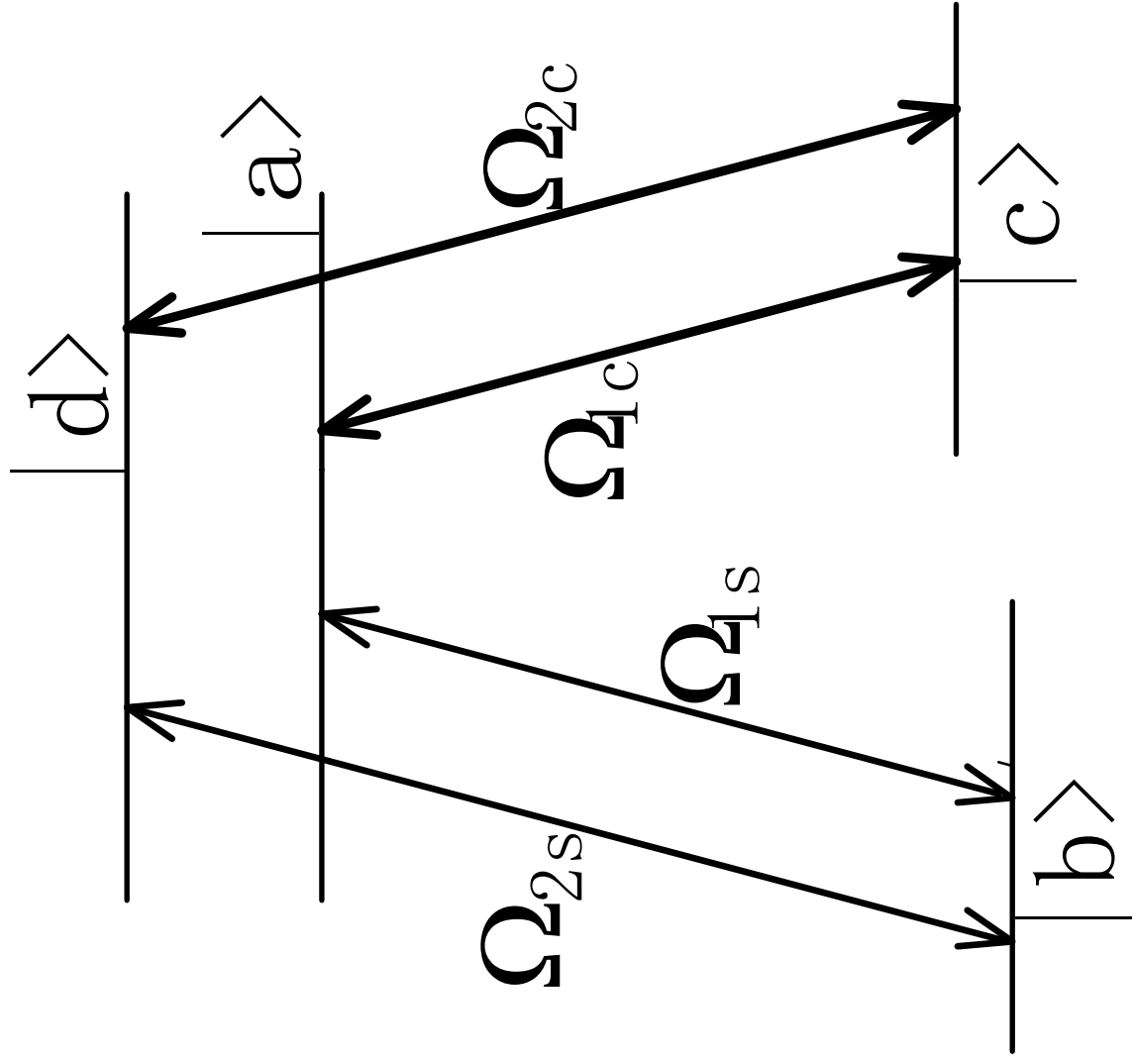
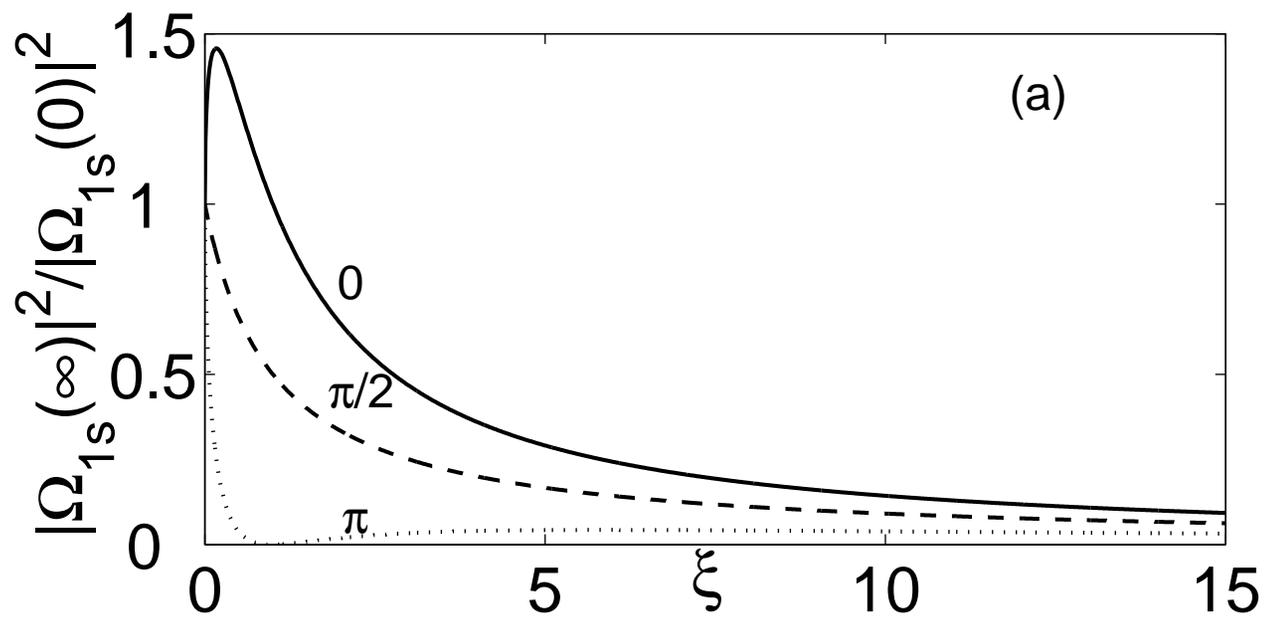
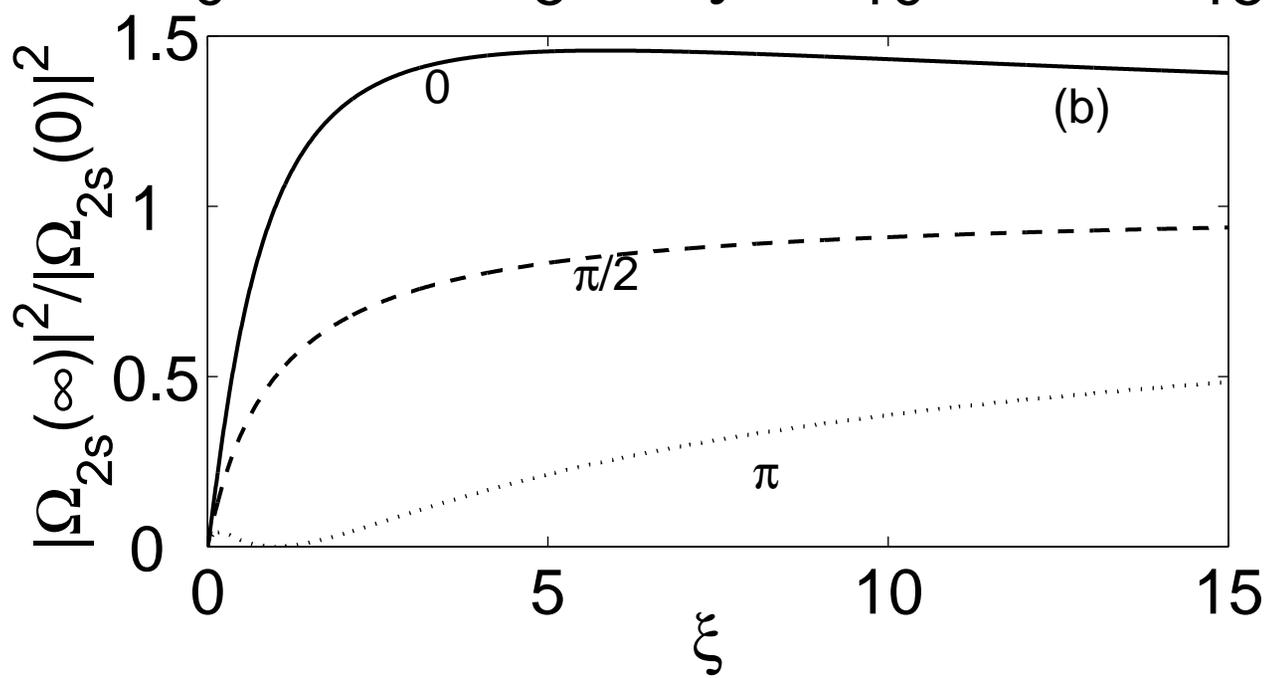
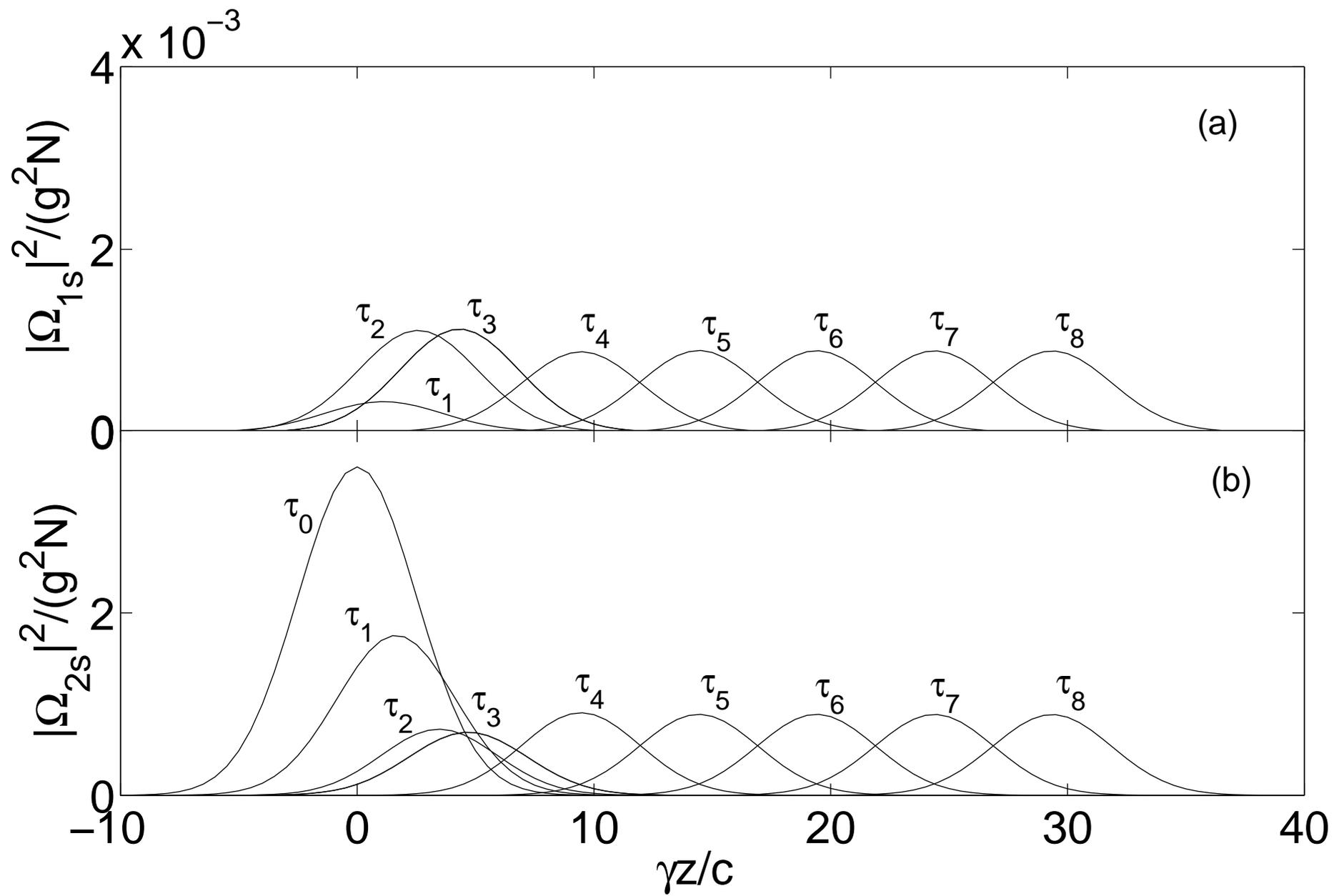
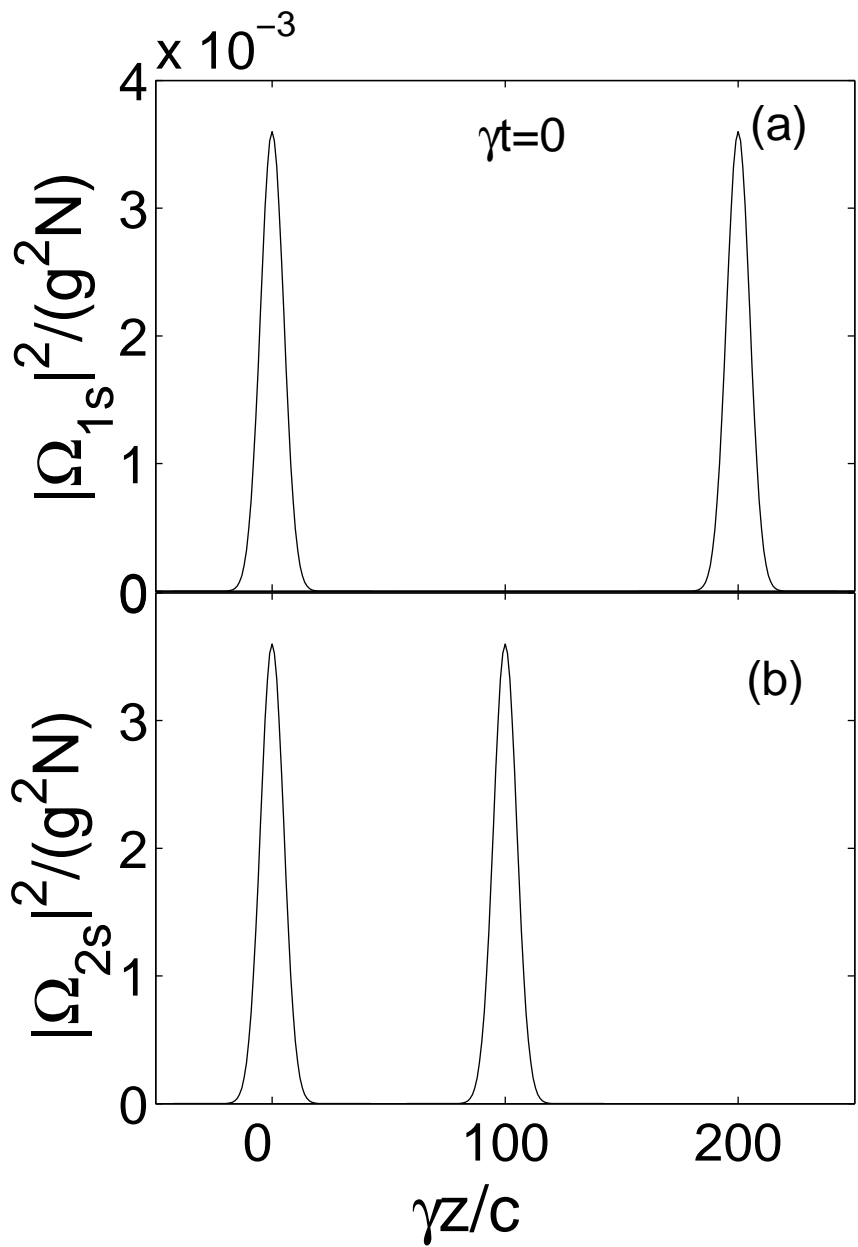
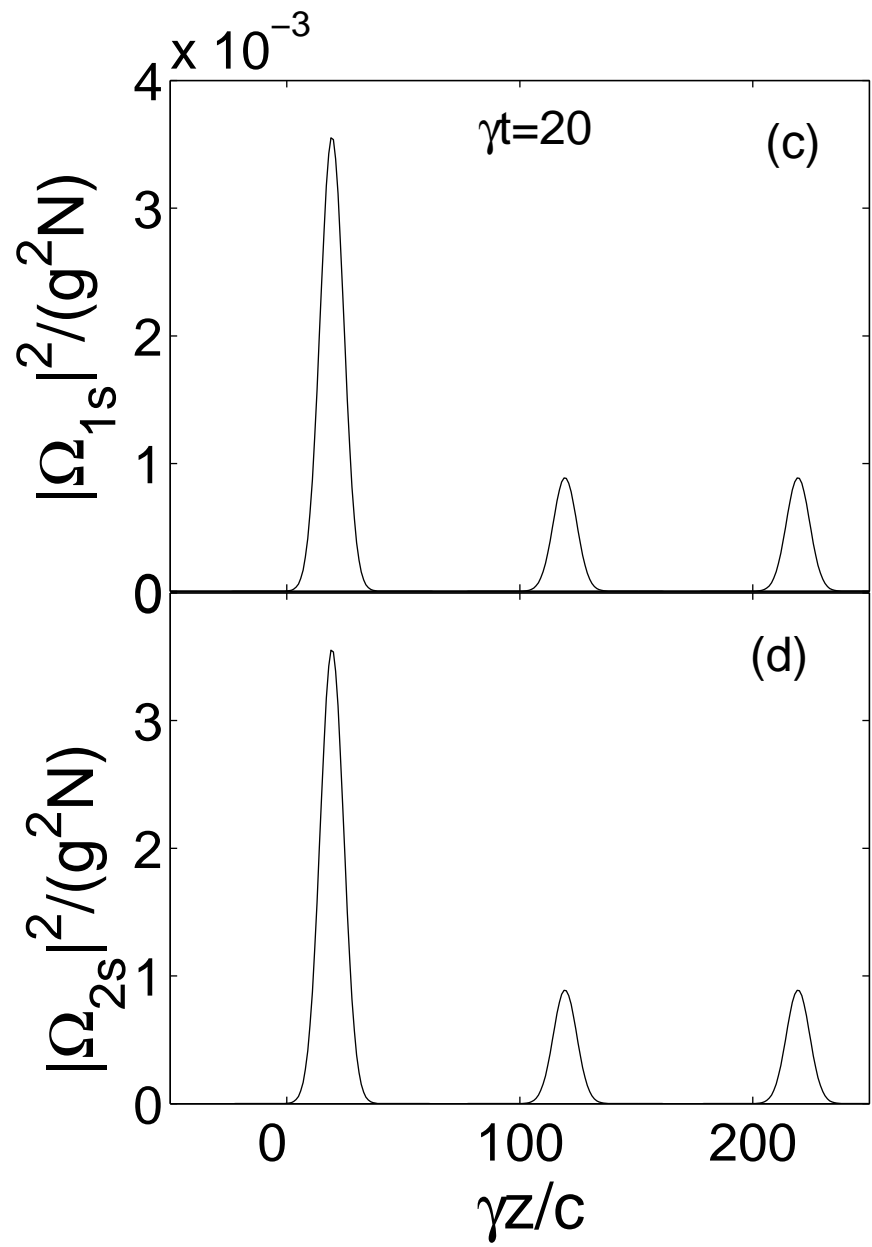
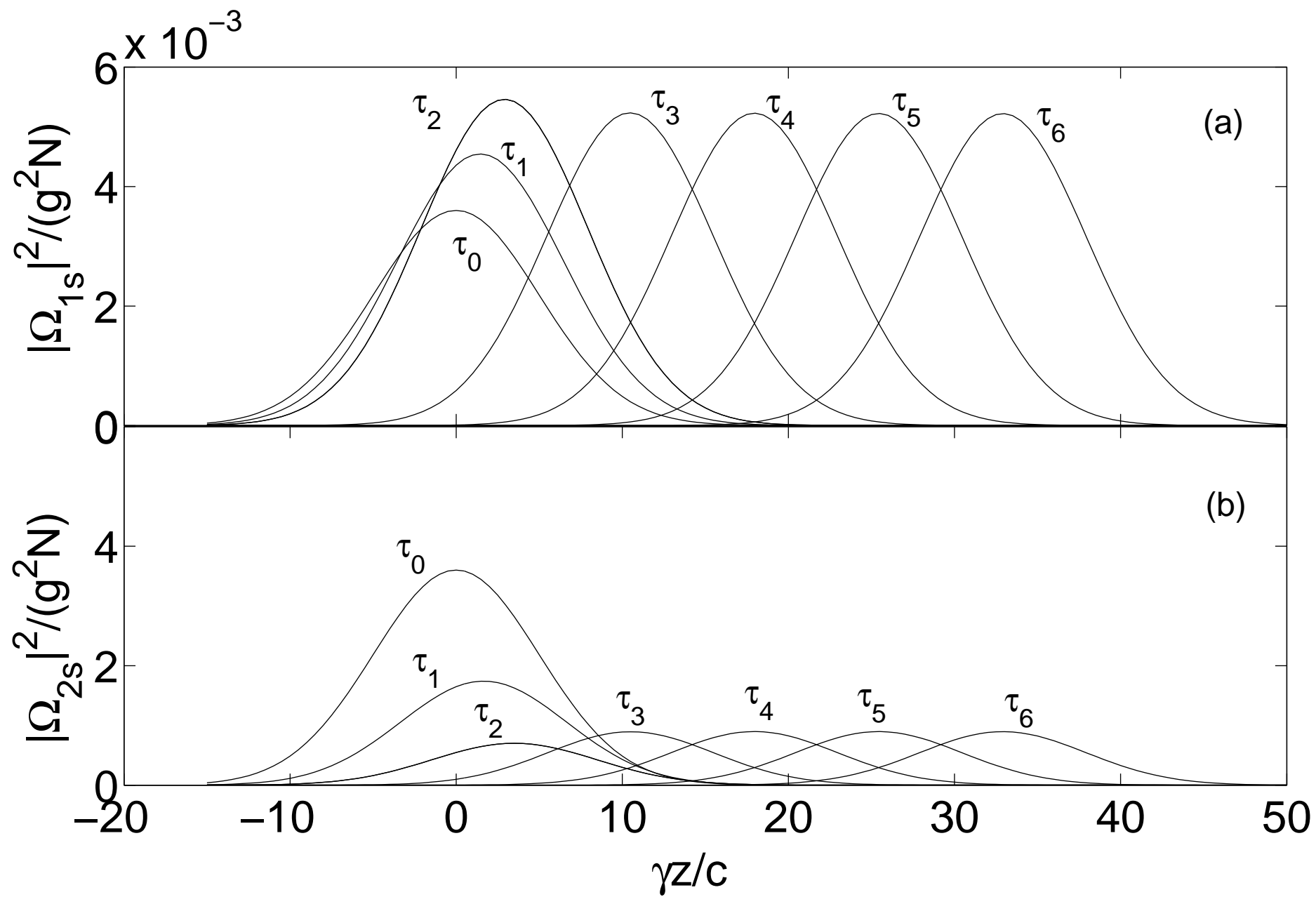
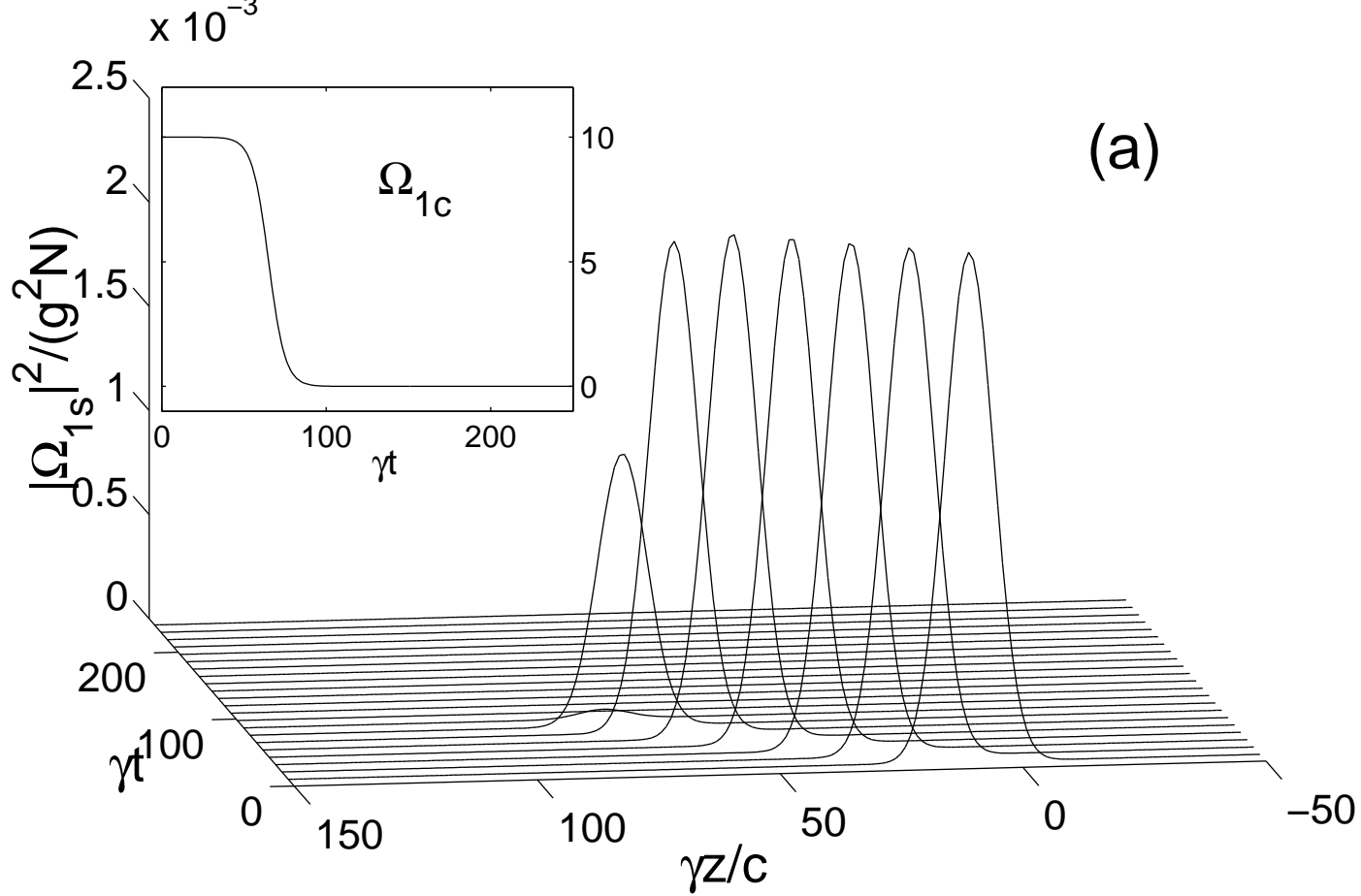
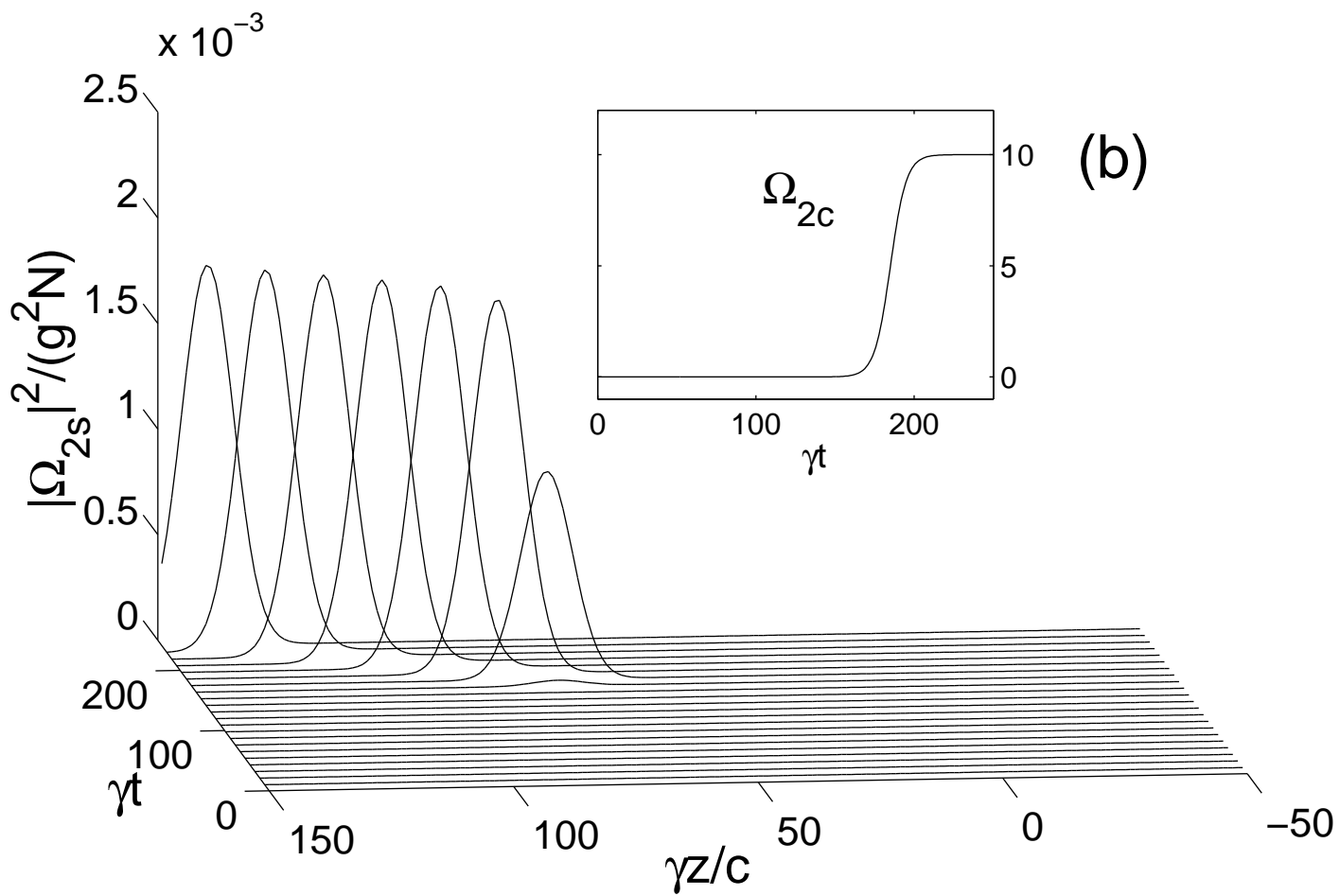
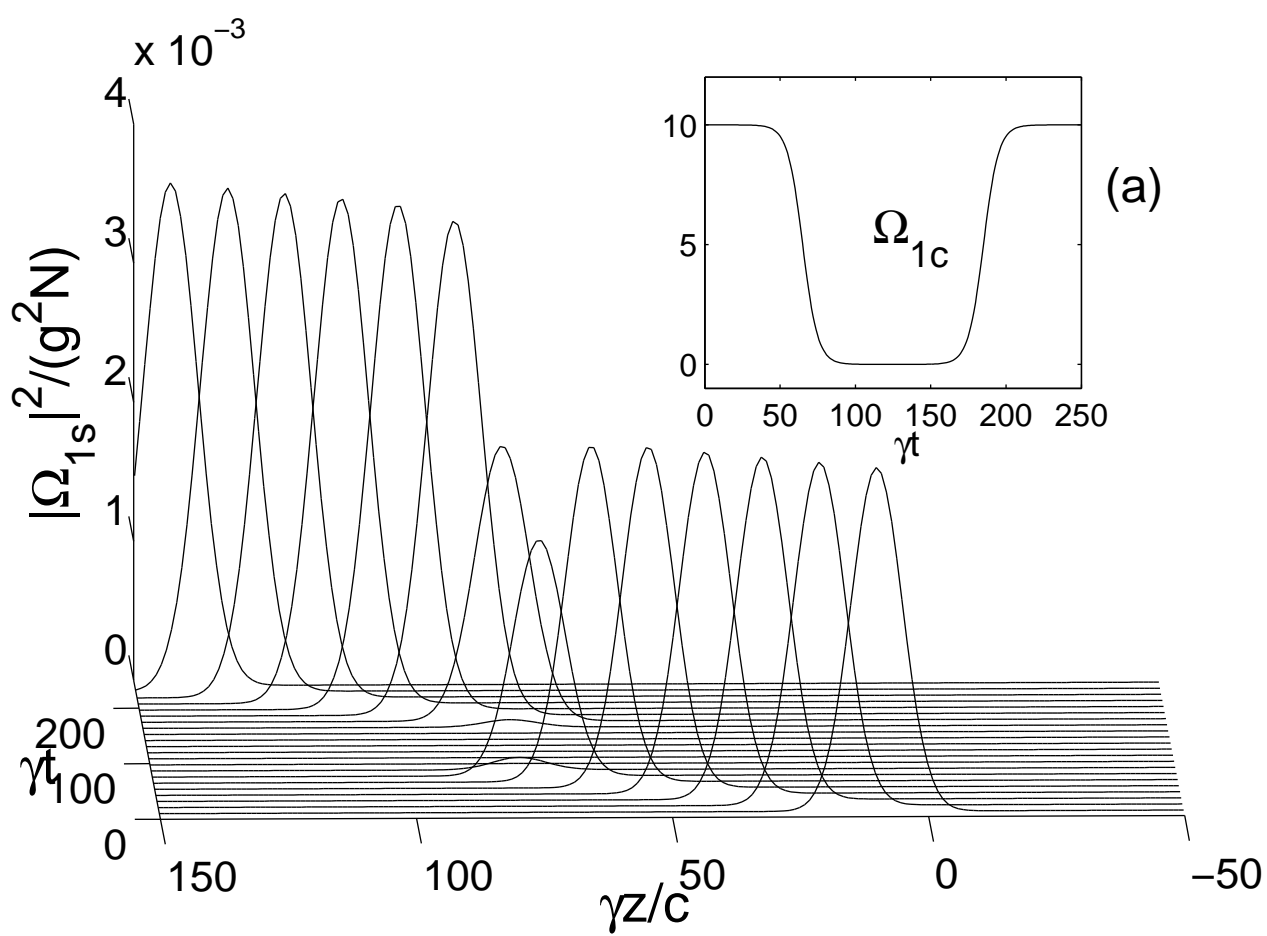
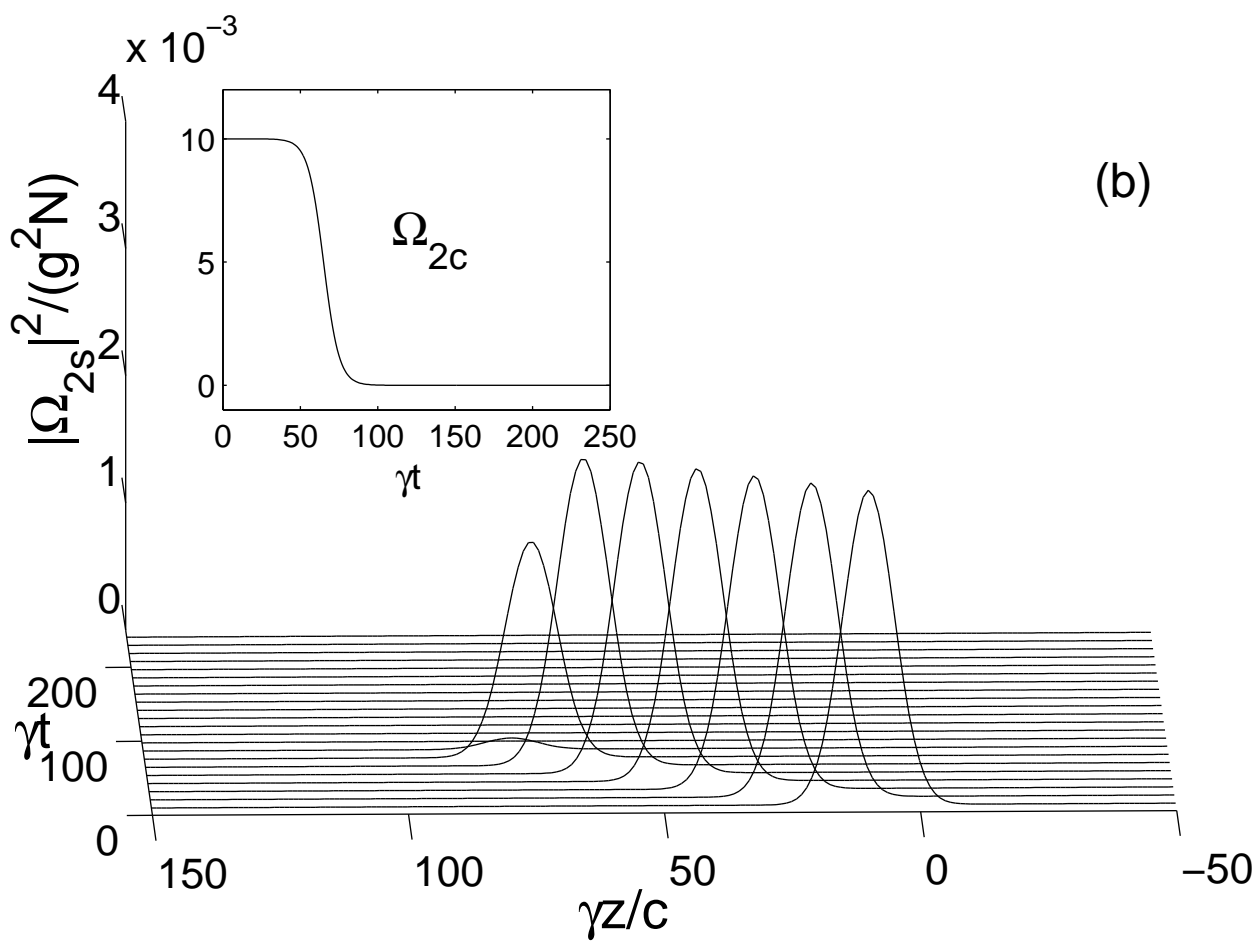
\end{document}